\documentclass[hidelinks,twocolumn,secnumarabic,amssymb, nobibnotes, aps, prb]{revtex4-2}
\usepackage{graphicx}
\usepackage{multirow}
\usepackage{lipsum} 
\usepackage{hyperref}
\usepackage[T1]{fontenc}
\usepackage[utf8]{inputenc}
\usepackage{textcomp} % for \textdegree
\graphicspath{{./figures/}}

\usepackage{xcolor}
\hypersetup{
    colorlinks,
    linkcolor={blue!80!black},
    citecolor={blue!80!black},
    urlcolor={blue!80!black}
}
\definecolor{hlcolor}{RGB}{209, 21, 7}
\newcommand{\hl}[1]{{\color{hlcolor} #1}}
\newcommand{\lozfept}{L1$_0$ FePt} 
\newcommand{\feptloz}{FePt L1$_0$}
\newcommand{\loz}{L1$_0$} 

\begin{document}

\title{\lozfept{} thin films with tilted and in-plane magnetic anisotropy: first-principles study}%

\author{Joanna Marciniak}%
\email[Corresponding author: ]{joanna.marciniak@ifmpan.poznan.pl}
\affiliation{Institute of Molecular Physics, Polish Academy of Sciences, M. Smoluchowskiego 17, 60-179 Poznań, Poland}
\date{\today}%
\author{Mirosław Werwiński}%
\affiliation{Institute of Molecular Physics, Polish Academy of Sciences, M. Smoluchowskiego 17, 60-179 Poznań, Poland}

\begin{abstract}
Ultrathin \lozfept{} films with different $c$-axis orientations relative to the film plane are promising candidates for data storage materials.
In this work, within the framework of density functional theory, we calculated the magnetic properties of ultrathin \lozfept{} (111) and (010) films with thicknesses ranging from 4 to 16 atomic monolayers (from about 0.8 to 3.5~nm). 
The highest average magnetic moments are observed for the thinnest films considered, and with increasing film thickness, the values converge towards the magnetic moment for bulk.
The observed increase comes mainly from enhanced moments in the two atomic monolayers closest to the surface of the films.
The easy axis of magnetization of (111) films prefers an alignment close to the tetragonal axis, an example of tilted magnetic anisotropy.
The 6-monolayer (111) film (about 1.3~nm thick) inclines the easy axis of magnetization of about 45° to the film plane, which can find use in applications.
The (010) films show an in-plane easy magnetization axis in a unique \loz{} tetragonal direction.
This is an unusual type of in-plane anisotropy, as the particular direction preference is very strong.
The computational results encourage further experimental studies of \loz{} systems with tilted and in-plane fixed magnetic anisotropy.

\end{abstract}

\maketitle

\section{Introduction}

The increase in data recording density is made possible by using new materials.
One of the approaches is using magnetic materials as thin films for this purpose because structuring can significantly improve the desired properties~\cite{siegel_synthesis_1993, gell_application_1995}.
One of the characteristic parameters defining the potential of a given ferromagnetic material for use in data storage media is magnetocrystalline anisotropy energy (MAE).
An example of a material with a very high positive MAE value, indicating strong uniaxial anisotropy, is the \lozfept{} phase.
The experimental MAE value of the \feptloz{} phase is about 11~MJ\,m$^{-3}$ at 4.2~K and 8~MJ\,m$^{-3}$ at 300~K, while its Curie temperature is about 750~K~\cite{hai_magnetic_2003}.
The density functional theory (DFT) calculations performed for the bulk phase show that the MAE can be in the range of 13.03~--~21.29~MJ\,m$^{-3}$ at 0~K~\cite{ayaz_khan_magnetocrystalline_2016, wolloch_influence_2017, marciniak_dft_2022} and that the spin magnetic moments are equal to 2.95~$\mu_\mathrm{B}$ on the Fe and 0.22~$\mu_\mathrm{B}$ on the Pt site.
The temperature dependence of the MAE in FePt can be used in heat-assisted magnetic recording (HAMR)-based memory devices, which can provide both density and speed improvements in data processing~\cite{christodoulides_copt_2000,thiele_ferhfept_2003,hovorka_curie_2012,weller_review_2016,john_magnetisation_2017}.
Magnetic data storage application of \lozfept{} films have been discussed in a review article by Lyubina~\textit{et al.}~\cite{lyubina_structure_2011}.

%-----------------on thin films (but not ultrathin)-----------------------------
%
Pioneering experiments using ordered tetragonal FePt magnetic thin films for data storage date back to the 1970s~\cite{evtikhiev_fept_1978}. 
But the most intensive research and technological development of methods for obtaining and analyzing FePt thin films for applications in spintronic devices occurred in the 1990s and the first two decades of the 21st century~\cite{lairson_epitaxial_1993, platt_l10_2002, yang_correction_2012}.
In most of the studies conducted, the FePt films have a thickness ranging from 10~nm to tens of~nm~\cite{lairson_epitaxial_1993, coffey_high_1995, shimatsu_thermal_1999, hsu_situ_2000, shima_preparation_2002, platt_l10_2002, takahashi_microstructure_2003, barmak_relationship_2005, you_particulate_2006, mihai_electron-magnon_2008, sallica_leva_magnetic_2010, matsumoto_magnetic_2011, yang_correction_2012}, in contrast to work focusing on ultrathin FePt films with thicknesses of single nanometers~\cite{coffey_high_1995, wu_low-temperature_2007, perumal_particulate_2008, kurth_finite-size_2010, kaidatzis_magnetic_2017, zhang_microstructure_2018}.
In the early 1990s, epitaxial thin films (about 50~nm thick) of FePt~(001) with perpendicular magnetic anisotropy were obtained~\cite{lairson_epitaxial_1993}.
In numerous cases, FePt films were deposited on a conventional MgO substrate~\cite{lairson_epitaxial_1993, takahashi_microstructure_2003, mihai_electron-magnon_2008, perumal_particulate_2008, kurth_finite-size_2010, zhang_microstructure_2018},
quite often on silver~\cite{hsu_situ_2000}, oxidized silicon~\cite{shimatsu_thermal_1999, sallica_leva_magnetic_2010, li_field-free_2022}, and other glassy substrates~\cite{sharma_l10_2011, wang_magnetization_2012}.
Experimentally obtained \lozfept{} thin films are usually not continuous but granular, which provides an additional degree of freedom that can be controlled to improve the desired properties and achieve dedicated patterned surface films~\cite{shima_preparation_2002}.
The main factor distinguishing FePt films from other materials is they possess one of the highest coercivity values~\cite{snarski-adamski_magnetic_2022}.
For example, Takahashi~\textit{et al.} obtained a huge coercivity of 42~kOe (4.2~T) for FePt films with a thickness of 10~nm~\cite{takahashi_microstructure_2003}.
In addition to the intrinsic properties of FePt, Takahashi~\textit{et al.} attributed this high value to the complete magnetic isolation of the nanoscale grains and the excellent uniaxial alignment of the $c$-axis perpendicular to the film plane~\cite{takahashi_microstructure_2003}.
Such a high coercive field value is comparable to those obtained for extremely strong samarium-cobalt permanent magnets.

An important factor affecting the magnetic properties of the compound is the degree of ordering of the L1$_0$ phase.
Optimal selection of technical parameters, such as the type and temperature of the substrate during deposition and the time and temperature of annealing, makes it possible to obtain FePt films with very high ordering parameter reaching 0.95~\cite{shima_preparation_2002, barmak_relationship_2005, yang_correction_2012}.

%-----------------about ultra thin films--------------------------------
%
The studies mentioned above of several nanometer-thick ultrathin FePt films reveal details of their granular microstructure~\cite{perumal_particulate_2008, kurth_finite-size_2010, kaidatzis_magnetic_2017, zhang_microstructure_2018}.
For example, for 2~nm thick films, the average grain size is about 8~nm, and the coercivity is about 17~kOe (1.7~T) with a degree of order ($S$) of about 0.6-0.7~\cite{perumal_particulate_2008}.
Highly ordered ($S=0.95$) highly textured granular films with grain diameters well below 5~nm achieve huge coercivity up to 7.3~T at room temperature~\cite{kurth_finite-size_2010}.
In contrast, reducing the nominal thickness to 0.5~nm decreases coercivity to 0.2~T~\cite{kurth_finite-size_2010}.
Zhang~\textit{et al.} for a FePt film thickness of 2.6~nm determined an averaged grain diameter of 4.3~nm~\cite{zhang_microstructure_2018} showing that the Curie temperature differs between the grains of different sizes and increase rapidly with decreasing grain diameter.
The magnetic anisotropy constant ($K\mathrm{_u}$) measured for 5~nm thick FePt film is equal to 4.2~MJ\,m$^{-3}$, while saturation magnetization is equal to 980~emu~cm$^{-3}$ (0.98~MA\,m$^{-1}$)~\cite{zhang_microstructure_2018}.

%------------------------------tilted magnetic anisotropy (TMA) ----------------------------%
Due to their tetragonal structure, \lozfept{} films are often obtained with a unique tetragonal $c$-axis perpendicular to the plane of the film.
This makes it possible to produce FePt films with perpendicular magnetic anisotropy that can be used in recording systems.
However, in 2005, Albrecht~\textit{et al.} experimentally obtained~\cite{albrecht_magnetic_2005} a system that, going beyond the standard configuration of longitudinal or perpendicular anisotropy, made it possible to achieve a magnetization direction tilted to the plane of the film, thus initiating a new path in the development of recording media~\cite{albrecht_magnetic_2005, wang_tilting_2005}.
At the same time, it was confirmed that tilting the magnetization direction to 45$^{\circ}$ minimizes the value of the switching field~\cite{albrecht_magnetic_2005}.
Since then, several other ways of achieving a film with tilted magnetization have been considered, such as ferromagnetic coupling of a tilted top Co layer with an in-plane magnetized bottom Co layer~\cite{lazarski_field-free_2019}. 
Much attention has been paid in particular to \lozfept{} and CoPt films~\cite{piskin_experimental_2021, li_field-free_2022}, which can be deposited on the substrate in such a way that the unique tetragonal axis is strongly deviated from the plane of the film, as in the case of the growth of the FePt films with a (111) surface~\cite{zha_pseudo_2009, zha_study_2009, sharma_l10_2011, wang_magnetization_2012}. 
The \lozfept{} (111) films in the aforementioned experiments have a thickness of 20 to 60~nm.
In contrast, the fabricated (111)-oriented CoPt single layer had a thickness of 5~nm~\cite{li_field-free_2022}.
The \lozfept{} (111) films were deposited on oxidized Si (001) and (100) surfaces~\cite{ristau_relationship_1999,sallica_leva_magnetic_2010}, Si (100) covered with 5~nm MgO~\cite{platt_l10_2002}, Ta~\cite{zha_pseudo_2009}, and metallic glasses such as CoFeTaB~\cite{sharma_l10_2011} and FeHfNbYB~\cite{wang_magnetization_2012}.
The measured magnetization direction of the \lozfept{}(111) film is tilted by 33-36° out-of-plane to the film plane~\cite{zha_pseudo_2009}, roughly in the tetragonal axis of the alternating Fe and Pt atomic monolayers.
FePt (111) films were also used in experimental pseudo spin valves~\cite{zha_pseudo_2009}, indicating a noticeable magnetoresistance, followed by a theoretical analysis of a spin valve with conical magnetization~\cite{matsumoto_spin-transfer-torque_2015}.
To directly measure angles of tilted magnetic anisotropy, Pişkin~\textit{et al.} proposed a new experimental method using anomalous and planar Hall effects together and called it angle-resolved Hall effect signals (ARHES)~\cite{piskin_experimental_2021}.

%---------------------------fixed in plane magnetic anisotropy----------------------%
Although the possibility of aligning the tetragonal crystallographic axis of the \lozfept{} film in a direction [111] inclined to the film plane is relatively well studied, the alignment of the tetragonal $c$ axis in the plane of the film appears less frequently in the literature~\cite{hsu_situ_2000,shima_preparation_2002,ohtake_l10_2012,sepehri-amin_microstructure_2017,wu_atomic-scale_2022}.
This alignment of the $c$-axis results in a magnetization oriented in a specific direction on the plane.
Such a uniaxial configuration could be called in-plane fixed magnetic anisotropy, as opposed to conventional in-plane magnetic anisotropy, when the approximately free rotation of the magnetization direction in the plane of the film is allowed.

%--------DFT papers on FePt films---------------------------------
%
All previous first-principles calculations of \lozfept{} films assumed a (001) surface (atomic Fe and Pt monolayers lying in the film plane)~\cite{zhang_electric-field_2009, lee_ferroelectric_2013, aas_exchange_2013, zhu_first-principles_2013, zhu_picosecond_2014, geranton_spin-orbit_2015, liu_modulation_2019, hammar_theoretical_2022}.
The FePt films considered were up to 19 atomic monolayers thick~\cite{hammar_theoretical_2022}.
The effect of film thickness on magnetic properties was investigated, and parameters such as surface magnetic anisotropy and bulk magnetic anisotropy were determined~\cite{zhang_electric-field_2009}.
In addition, the properties of FePt films coated with Fe and Pt layers~\cite{zhang_electric-field_2009, aas_exchange_2013, geranton_spin-orbit_2015} and FePt films on MgO substrate with Cu and Pt overlays~\cite{zhu_first-principles_2013, zhu_picosecond_2014} were also studied from first-principles.
The hard magnetic properties of large FePt grains were also modeled~\cite{antoniak_guideline_2011}, and the possibility of ferroelectric control of magnetic anisotropy in FePt/BaTiO$_3$~\cite{lee_ferroelectric_2013} and FePt/PbTiO$_3$~\cite{liu_modulation_2019} heterostructures were computationally analyzed.

%----------summary of what will be presented-----------
%
In the present work, we focus on the first-principles study of \lozfept{} thin films with (111) and (010) surfaces, limiting ourselves to freestanding FePt films surrounded on the top and bottom sides by vacuum (without including the substrate material and overlayer).

\section{Calculations' details}

\begin{figure}[t]
\centering
\includegraphics[width=0.97\columnwidth]{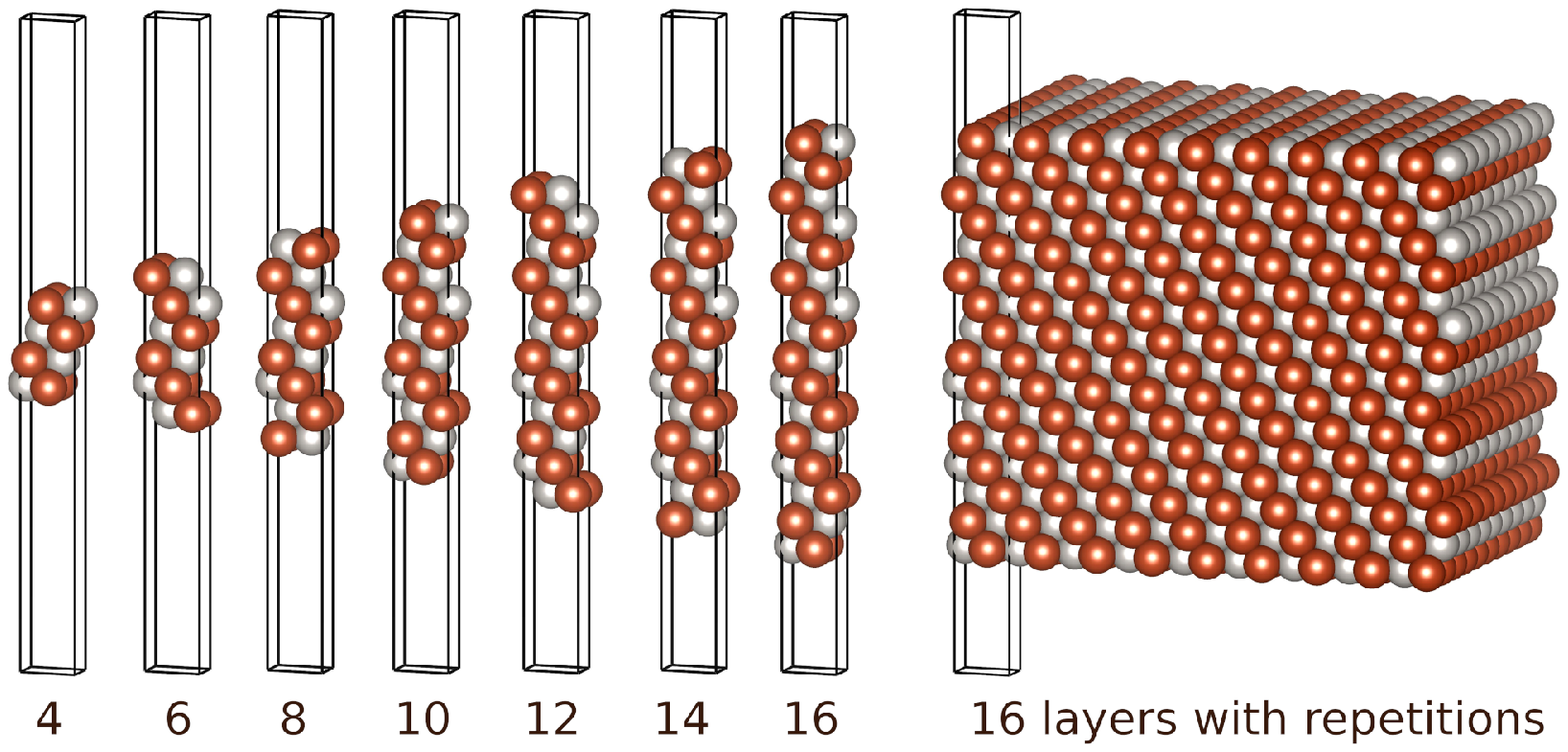}
\vspace{5mm}
(a) \lozfept{} (111) thin films

\includegraphics[width=0.97\columnwidth]{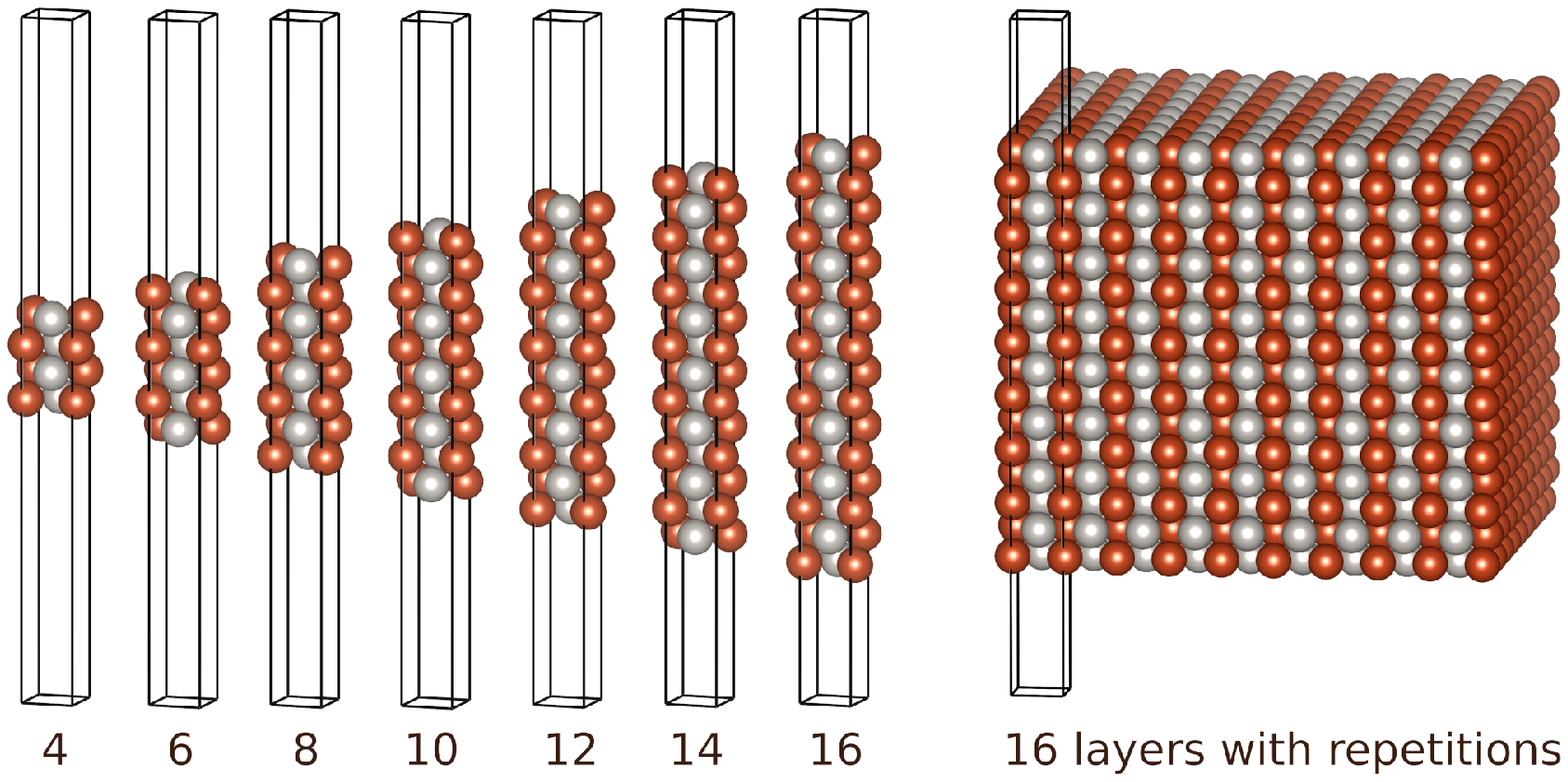}
(b) \lozfept{} (010) thin films
\caption{\label{fig-111-structure}
Compilation of unit cells of \lozfept{} thin films with (111) and (010) surfaces and thicknesses ranging from 4 to 16 atomic monolayers (from about 0.9 to 3.5~nm for (111) films and from 0.78 to 3.1~nm for (010) films). 
On the right, fragments of infinite films are generated by duplicating 16-monolayer unit cells several times in directions in the film plane.
}
\end{figure}

\begin{table}[]
\caption{Structural data of \lozfept{} thin films with 16 atomic monolayers and surfaces (111) and (010).
The unit cell parameters $c$ specify the height of the computational unit cell, including the vacuum.
The thickness of the 16-monolayer FePt (010) film is 3.1~nm
The thicknesses of the (111) and (010) films themselves are 35.47 and 30.99~\AA{}, respectively.
}
\label{tab-wyckoff-pos}
\begin{tabular}{c | c c c | c c c}
        \hline
        \hline
        \multirow{4}{*}{Element} & \multicolumn{3}{c}{(111) thin film} & \multicolumn{3}{|c}{(010) thin film}\\
         & \multicolumn{3}{c}{$P 1 2/m 1$}     &\multicolumn{3}{|c}{$P m m a$} \\
        & $a$ (\AA) & $b$ (\AA) & $c$ (\AA)      & $a$ (\AA) & $b$ (\AA) & $c$ (\AA) \\
            &  2.738    &  4.654    & 53.638      &  3.872    &  3.763    & 48.914\\
        \hline
        Fe    &  0.000    & -0.173    & -0.479    &  0.250    &  0.000    &  0.480\\
        Pt    &  0.500    &  0.327    & -0.479    & -0.250    &  0.500    &  0.480\\
        Fe    &  0.000    &  0.481    & -0.438    & -0.250    &  0.000    &  0.441\\
        Pt    & -0.500    & -0.019    & -0.438    &  0.250    &  0.500    &  0.441\\
        Fe    &  0.000    &  0.135    & -0.397    &  0.250    &  0.000    &  0.401\\
        Pt    &  0.500    & -0.365    & -0.397    & -0.250    &  0.500    &  0.401\\
        Fe    &  0.000    & -0.212    & -0.356    & -0.250    &  0.000    &  0.361\\
        Pt    & -0.500    &  0.288    & -0.356    &  0.250    &  0.500    &  0.361\\
        Fe    &  0.000    &  0.442    & -0.314    &  0.250    &  0.000    &  0.322\\
        Pt    &  0.500    & -0.058    & -0.314    & -0.250    &  0.500    &  0.322\\
        Fe    &  0.000    &  0.096    & -0.273    & -0.250    &  0.000    &  0.282\\
        Pt    & -0.500    & -0.404    & -0.273    &  0.250    &  0.500    &  0.282\\
        Fe    &  0.000    & -0.250    & -0.232    &  0.250    &  0.000    &  0.243\\
        Pt    &  0.500    &  0.250    & -0.232    & -0.250    &  0.500    &  0.243\\
        Fe    &  0.000    &  0.404    & -0.190    & -0.250    &  0.000    &  0.203\\
        Pt    & -0.500    & -0.096    & -0.190    &  0.250    &  0.500    &  0.203\\
        \hline
        \hline
    \end{tabular}
\end{table}

%-----------------structur models-----------------------------------
%
The starting point for the preparation of thin film models was the bulk phase crystallographic unit cell of \lozfept{} (space group $P$4$/mmm$, $a$'~=~3.872/$\sqrt{2}$~=~2.738~\AA{} and $c$~=~3.763~\AA{}), which we optimized earlier~\cite{marciniak_dft_2022}.
Based on the tetragonal unit cell, two sets of films ranging from 4 to 16 atomic monolayers were generated with (111) and  (010) surfaces, see Fig.~\ref{fig-111-structure}.
The complete structural data of the \lozfept{} films with (111) and (010) surfaces with the extreme thicknesses of the 16 atomic monolayers are presented in Table~\ref{tab-wyckoff-pos}.
The vacuum space in the direction normal to the film is added so that films of different thicknesses have a constant value of the unit cell lattice parameter $c$.
The minimum height of vacuum space in the unit cell is about 22~\AA{} in 16-monolayer systems.
The film models with smaller thicknesses were obtained by subtracting the outer monolayers.
None of the structures considered were subjected to film geometry optimization.
Such a simplification can be justified by results from Hammar~\textit{et al.}~\cite{hammar_theoretical_2022}, where optimization of the geometry of FePt film showed that mainly the positions of only three outer atomic monolayers change. The change in the distance between these layers is only about 1~--~2\% and does not significantly affect the magnetic properties.
All structure representations were prepared using VESTA code~\cite{momma_vesta_2008}.

%---------------------DFT calculations---------------------------------------------
%
First-principles calculations were performed using the full-potential local-orbital code (FPLO18.00-52) developed to solve the Kohn-Sham equations of density functional theory~\cite{koepernik_full-potential_1999, eschrig_chapter_2004}.
The Perdew-Burke-Ernzerhof exchange-correlation potential (PBE)~\cite{perdew_generalized_1996} was used.
For each model, self-consistent scalar-relativistic calculations were performed with a $60 \times 30 \times 5$ \textbf{k}-point mesh, a density convergence criterion of 10$^{-6}$, and the tetrahedron method chosen for integration over the Brillouin zone.
Finally, to find the magnetic anisotropy energy, one 
fully-relativistic iteration~\cite{eschrig_chapter_2004} was performed for each of the selected quantization directions of the considered films.

%------------determination of directions of magnetization------------------
%
%
The calculations of the complete energy dependence of the systems on the magnetization direction were carried out for 10-monolayer structures with surfaces (111) and (010).
To remove energy discrepancies between different magnetization directions due to variation in the choice of \textbf{k}-mesh symmetry, these particular calculations were performed using computational unit cells with the space group reduced to \textit{P}1.
The magnetization direction was determined for a set of films with a surface (111) in the following way.
Having found that the energy minimum must be in the plane (100) for each film thickness, calculations were made for directions in this plane varying every 2$^{\circ}$.
Based on five such results, the direction of the easy axis of magnetization was determined using a third-degree polynomial fit.

\begin{figure*}[]
\centering
\includegraphics[width=0.92\textwidth]{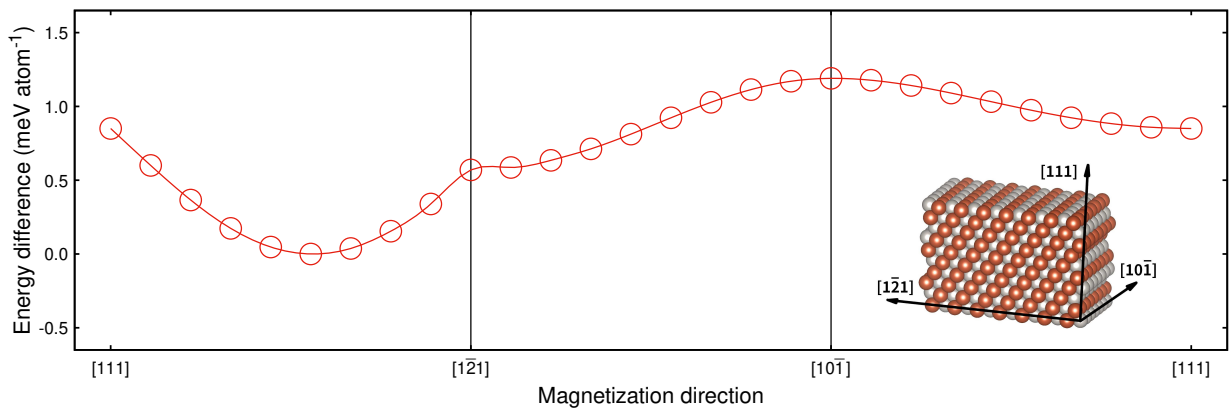}
\caption{\label{fig-en-dir-111}
Evolution of magnetic anisotropy energy with change in magnetization direction for a 10-monolayer \lozfept{} film with (111) surface.
The energy difference is determined between the total energies for a given direction and the direction of the axis of easy magnetization (energy minimum).
The energy minimum occurs at an angle of about 50.0$^{\circ}$ relative to the \hl{[111]} direction.
Calculations were performed with the PBE exchange-correlation potential using the FPLO18 code.
}
\end{figure*}

%----------weaknesses of the model------------------------------
%
Below, we discuss some of the approximations adopted in this work, resulting from the need to reduce the computational complexity of the systems.
These approximations can be divided into groups concerning crystallographic and electronic structure.
%
%--------infinite films - no edges, no shape anisotropy
%
The first group involves the assumption of infinite periodicity of the model in the film plane.

This leads to a failure to incorporate finite layer and grain sizes of real materials in the magnetic shape anisotropy.
%
%--------no substrate and no overlay 
%--------no Pt layer on the overlay
%
Furthermore, magnetic films are usually deposited on substrates and coated with a cap layer in practical applications. 
Interactions of the magnetic film with the substrate and overlayer can significantly affect the properties of the system.
However, our models do not consider the substrate or overlayer.
This assumption was due to the need to constrain the models to allow accurate, fully-relativistic calculations to determine magnetic anisotropy energies.
%
%--------perfect L10 order (100%)
%
Another assumption is the fully ordered arrangement of atoms. 
Therefore, defects such as atom swapping or extra atoms in interstitial positions are not accounted for. 
%
%--------no grains, single phase
%--------atomically flat surface
%
Additionally, the models assume no other phases, no grains, and a perfect, atomically smooth surface of the films.
Subsequently, it was also necessary to use approximations within the electronic structure.
Several of these have been mentioned previously, but we will bring up two more.
%
%--------PBE
%--------0K temperature
%
DFT calculations were performed for the ground state, meaning that the films' properties were determined for the temperature of 0~K.
The obtained results of such quantities as magnetic moments and magnetic anisotropy may, therefore, differ from those observed at room temperature.
%
%-------collinear magnetic configuration-------------------
%
In addition, we also assumed a collinear arrangement of magnetic moments, which may not be well met for the film surfaces.

%--------------conclusion on weaknessess----------------------
%
Despite the series of approximations summarized above, the results obtained for \lozfept{} (111) and (010) films represent a significant step toward a better understanding of these structured materials. 
The presented calculations also represent the first attempt to model \lozfept{} films with an arrangement of atomic monolayers other than in the plane of the films. 
They are one of the first to study the thickness dependence of the magnetic properties of FePt films.

\section{Results and discussion}

Among the magnetic compounds with the \loz{} layered structure, FePt, FePd, FeNi, and CoPt are of particular interest.
In particular, their potential use for data storage makes them intensively studied, especially in the form of thin films.
The thickness of experimentally investigated \loz{} films is sometimes less than 1~nm (about five atomic monolayers)~\cite{kurth_finite-size_2010}.
Several experimental studies also cover the range of ultrathin FePt films with thicknesses on the order of single nanometers.
Among \lozfept{} films, systems with Fe and Pt monolayers arranged in the [001] direction are the most commonly studied.
However, in the second decade of the 21st century, a lot of attention has also been paid by experimenters to \lozfept{} (111) films due to the observed tilted magnetic anisotropy and much less to \lozfept{} (010) films with the direction of alternating Fe/Pt monolayers perpendicular to the film plane.
These two types of surfaces are also important from a practical point of view because, as previous calculations have shown, the energetically most stable \lozfept{} surface is (111), followed by (100/010), and only the third most stable is (001)~\cite{dannenberg_surface_2009}.
So far, however, no computational analysis of the \lozfept{} (111) and (010) films has been performed under first-principles calculations.
Since we are convinced this knowledge gap is worth filling, below we will present the computational results for ultrathin films (111) and (010) with thicknesses ranging from about 0.8 to 3.5~nm.

\subsection{Thin films of \lozfept{} (111)}

\begin{figure}[t]
\centering
\includegraphics[clip,width=0.92\columnwidth]{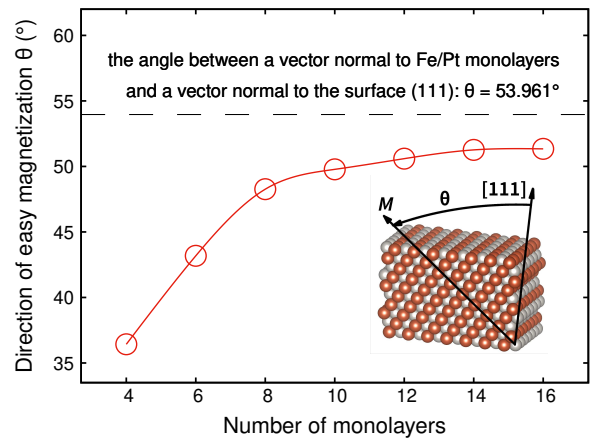}
\caption{\label{fig-dir-thickness}
The easy magnetization axis direction($\theta$) as a function of the thickness of \lozfept{} films with a surface (111).
The angle between the vector normal to the Fe/Pt monolayers and the vector normal to the film surface (53.961$^{\circ}$) is indicated as a dashed line.
Calculations were performed with the PBE exchange-correlation potential using the FPLO18 code.}
\end{figure}

\begin{figure}[t]
\includegraphics[clip,width=0.98\columnwidth]{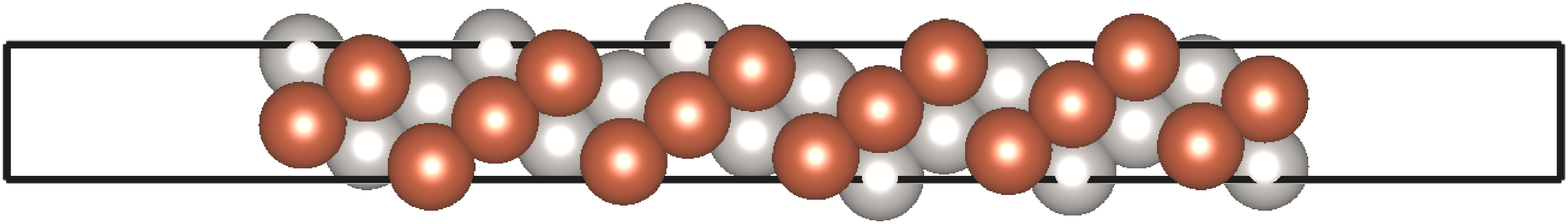}
\includegraphics[clip,width=0.98\columnwidth]{fept_16_scalar_excess_e_vs_position.eps}
\caption{\label{fig-m_e_vs_position_111} Charge transfer (a) and spin magnetic moment (b) in a 16-monolayer (111) film of \lozfept{}, together with the unit cell shown on top as a function of atomic position in a direction perpendicular to the film plane.
Calculations were performed with the PBE exchange-correlation potential in scalar-relativistic formalism using the FPLO18 code.}
\end{figure}

\begin{figure*}[t]
\includegraphics[clip,width=0.92\textwidth]{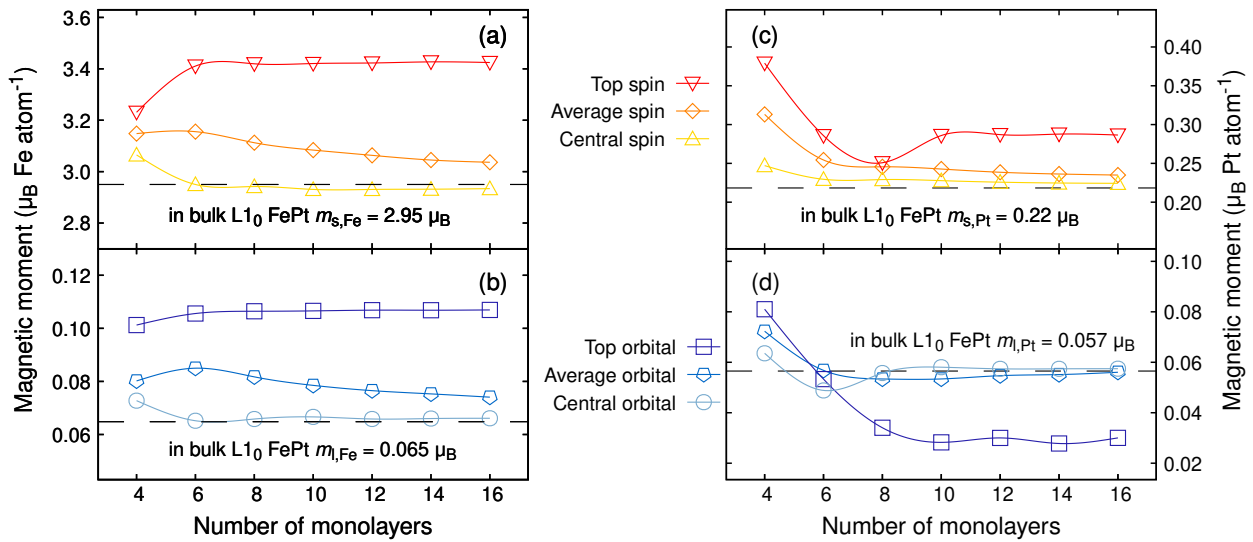}
\caption{\label{fig-mag-mom-atoms-111}
Spin and orbital magnetic moments at Fe (a, b) and Pt (c, d) atoms as a function of the number of monolayers in \lozfept{} films with (111) surface obtained for easy magnetization direction.
Magnetic moments at specified atoms for bulk \lozfept{} system are indicated as dashed horizontal lines.
The calculations were performed with the PBE exchange-correlation potential using the FPLO18 code.}
\end{figure*}

\begin{figure}[t]
\includegraphics[clip,width=0.9\columnwidth]{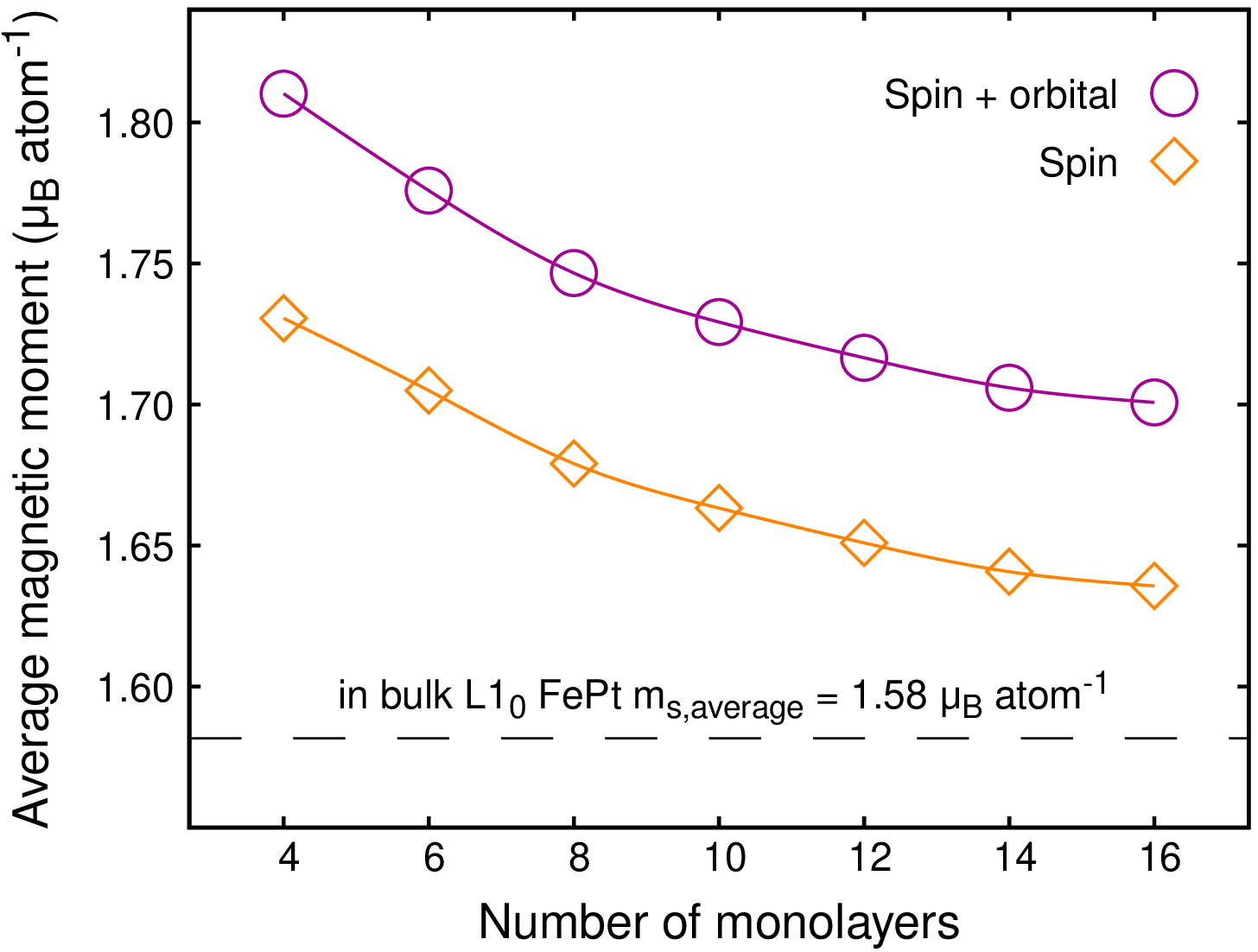}
\caption{\label{fig-mag-mom-dependence_111}
Average total (spin + orbital) and spin magnetic moments for easy magnetization direction \textit{versus} a number of atomic monolayers in \lozfept{} film with (111) surface.
A dashed horizontal line indicates the average spin magnetic moment for a bulk \lozfept{} phase.
 The calculations were performed with the PBE exchange-correlation potential using the FPLO18 code.}
\end{figure}

%--------introduction to 111 films-----------------
%
We will begin by discussing the results obtained for films (111) as systems already better-recognized experimentally~\cite{zha_pseudo_2009, zha_study_2009, sharma_l10_2011, wang_magnetization_2012}.
We will present computational results on how magnetic moments and magnetization direction change with the film thickness.

%------------structural models of (111) films-----------------------------
%
Figure~\ref{fig-111-structure}(a) shows a set of prepared unit cells of \lozfept{} (111) with thicknesses ranging from 4 to 16 atomic monolayers. 
Since the models are infinitely periodic, the shown unit cells describe films infinite in a plane. 
The thickness of the 16-monolayer film is 3.5~nm, corresponding to a distance of 0.22~nm between atomic monolayers in the [111] direction.
Thus, the thinnest 4-monolayer (111) film's thickness is 0.88~nm.

%----------------magnetic anisotropy in 10-monolayers film----------------%
%
We begin the analysis of \lozfept{} films with (111) surface by determining the total energy dependence of the system as a function of magnetization direction for a film with a thickness of 10 atomic monolayers, see Fig.~\ref{fig-en-dir-111}.
Calculations were made for magnetization directions changing every 10$^{\circ}$ along the path [$111$]~$\to$~[$1\bar{2}1$]~$\to$~[$10\bar{1}$]~$\to$~[111].
The selected 10-monolayer film shows an energy minimum at an angle of about 50$^{\circ}$ to the [111] direction. 
This means that the determined direction of magnetization is not precisely parallel to the direction of alignment of the Fe and Pt monolayers, which is equal to 53.96$^{\circ}$ (counting from the [111] direction, not from the plane of the film, as done in Ref.~\cite{zha_pseudo_2009}).
The energy difference between the easy magnetization direction and the direction normal to the films is about 0.85~meV\,atom$^{-1}$ ($\sim$10~MJ\,m$^{-3}$), which is smaller than the 1.38~meV\,atom$^{-1}$ for bulk FePt~\cite{marciniak_dft_2022}, but still implies very high magnetic anisotropy, as seen in Fig.~\ref{fig-en-dir-111}.

%---------------magnetization direction versus film thicknesss----------------%
%
Figure~\ref{fig-dir-thickness} shows the dependence of the easy magnetization direction on film thickness.
For larger thicknesses, the magnetization angle converges in the direction of the alignment of Fe and Pt monolayers ($\theta = 53.96^{\circ}$).
However, for thinner films, the direction of magnetization deviates in the out-of-plane direction.
For films with a thickness of six atomic monolayers (1.3~nm), the angle of easy magnetization is about 45$^{\circ}$, which is the direction that minimizes the value of the switching field in systems with tilted magnetic anisotropy~\cite{albrecht_magnetic_2005, wang_tilting_2005}.
However, magnetically hard FePt films alone are not best suited for switching. 
Instead, it is conceivable that a magnetically hard film would be coupled to a magnetically soft layer, which can be switched much more easily.

%--------------charge transfer and magnetic moment in 16-monolayers film------------%
%
Figure~\ref{fig-m_e_vs_position_111} shows two more characteristics relatively universal for the entire set of (111) films under consideration: the distributions of charge transfer and spin magnetic moment in an example film with a thickness of 16 monolayers.
The sample's unit cell parameter $c$ equals 53.6~\AA{}, and the film thickness is 35.4~\AA{}.
Since we chose to make the $x$ axis centered at zero, the scale extends by 26.8~\AA{} in both directions.
The charge transfer plot reveals significant deviations for the three monolayers near the surface, and electron deficiency on Fe atoms is compensated for on Pt atoms, see Fig.~\ref{fig-m_e_vs_position_111}(a).
Like the excess electrons, the spin magnetic moments of several monolayers near the surface stand out clearly.
Moreover, the spin magnetic moments induced on Pt are proportional to those on Fe.
The central and surface spin magnetic moments on Fe atoms are significantly elevated relative to the value for bcc Fe of 2.15~$\mu_\mathrm{B}$~\cite{moruzzi_ferromagnetic_1986}.

%-----------magnetic moments------------------------------
%
Figures~\ref{fig-mag-mom-atoms-111} and \ref{fig-mag-mom-dependence_111} show the dependence of component and average magnetic moments on film thickness.
The fully-relativistic values of the spin (orbital) magnetic moment on the surface of the 16-monolayer film are 3.42 (0.11) and 0.29 (0.03)~$\mu_\mathrm{B}$\,atom$^{-1}$ for Fe and Pt atoms, respectively.
They are elevated compared to the bulk \lozfept{} values, for which the spin (orbital) magnetic moments are equal to 2.95 (0.065) and 0.22 (0.057)~$\mu_\mathrm{B}$\,atom$^{-1}$ for Fe and Pt atoms, respectively~\cite{marciniak_dft_2022}.
On the other hand, as we can see in Fig.~\ref{fig-mag-mom-atoms-111}, the values of magnetic moments we observe in the center of the film are comparable to the bulk values.
However, additional noticeable changes in the values of magnetic moments are observed for the thinnest films with a thickness of fewer than 10 monolayers.
Particularly for the 4-monolayer film, we observe a significant increase in the spin and orbital magnetic moments on Pt atoms on the surface.
The average spin (total) magnetic moment of 1.58 (1.64)~$\mu_\mathrm{B}$\,atom$^{-1}$  is also elevated relative to the bulk values and increases further with decreasing film thickness.
This is due to elevated magnetic moments on the film surface, whose contribution to the average becomes more significant with decreasing film thickness.
An analogous dependence of the magnetic moment on the layer thickness was observed for ultrathin Fe films, both experimentally and theoretically~\cite{beltran_interfacial_2012,scheunert_review_2016}.

%-------------conclusions for (111) films------------------------
%
In summary, we have shown that the ultrathin \lozfept{} (111) films are characterized by high-valued and tilted magnetic anisotropy and relatively high values of magnetic moments.
Such strongly tilted magnetic anisotropies can find use, for example, as hard magnetic films for coupling with soft magnetic films, inducing tilted magnetization in the latter.

\subsection{Thin films of \feptloz{} (010)}

\begin{figure*}[t]
\includegraphics[width=0.92\textwidth]{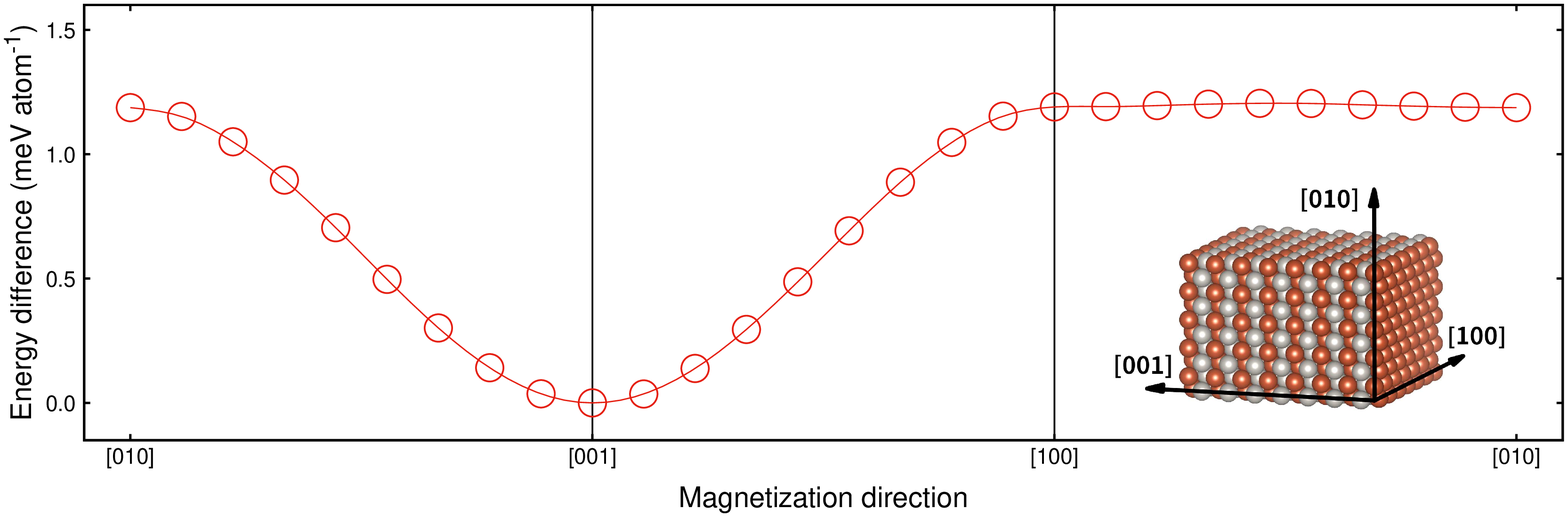}
\caption{\label{fig-en-dir-010}
Evolution of magnetic anisotropy energy with change in magnetization direction for a 10-monolayer film of \lozfept{} with (010) surface. 
The energy difference is determined between the total energies for a given direction and the direction of the axis of easy magnetization (energy minimum).
Calculations were performed with PBE exchange-correlation potential using the FPLO18 code.
}
\end{figure*}

\begin{figure}[t]
\includegraphics[clip,width=0.92\columnwidth]{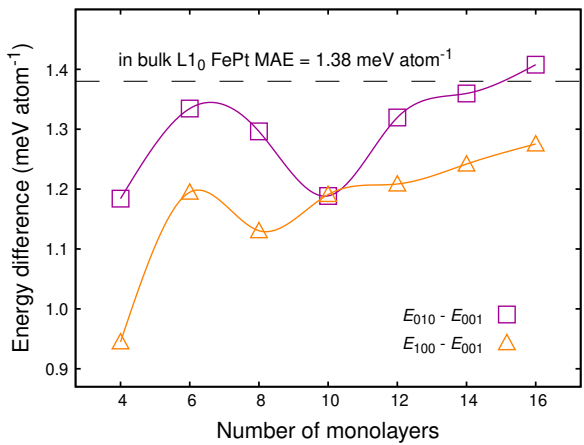}
\caption{\label{fig-en-diff-dependence}
Energy differences $E_{010}-E_{001}$ and $E_{100}-E_{001}$ as a function of thickness of \lozfept{} films with (010) surface.
The energies $E_{001}$, $E_{010}$, and $E_{100}$ were determined for the corresponding magnetization directions using the FPLO18 with the PBE exchange-correlation potential.}
\end{figure}

\begin{figure}[t]
\includegraphics[clip,width=0.92\columnwidth]{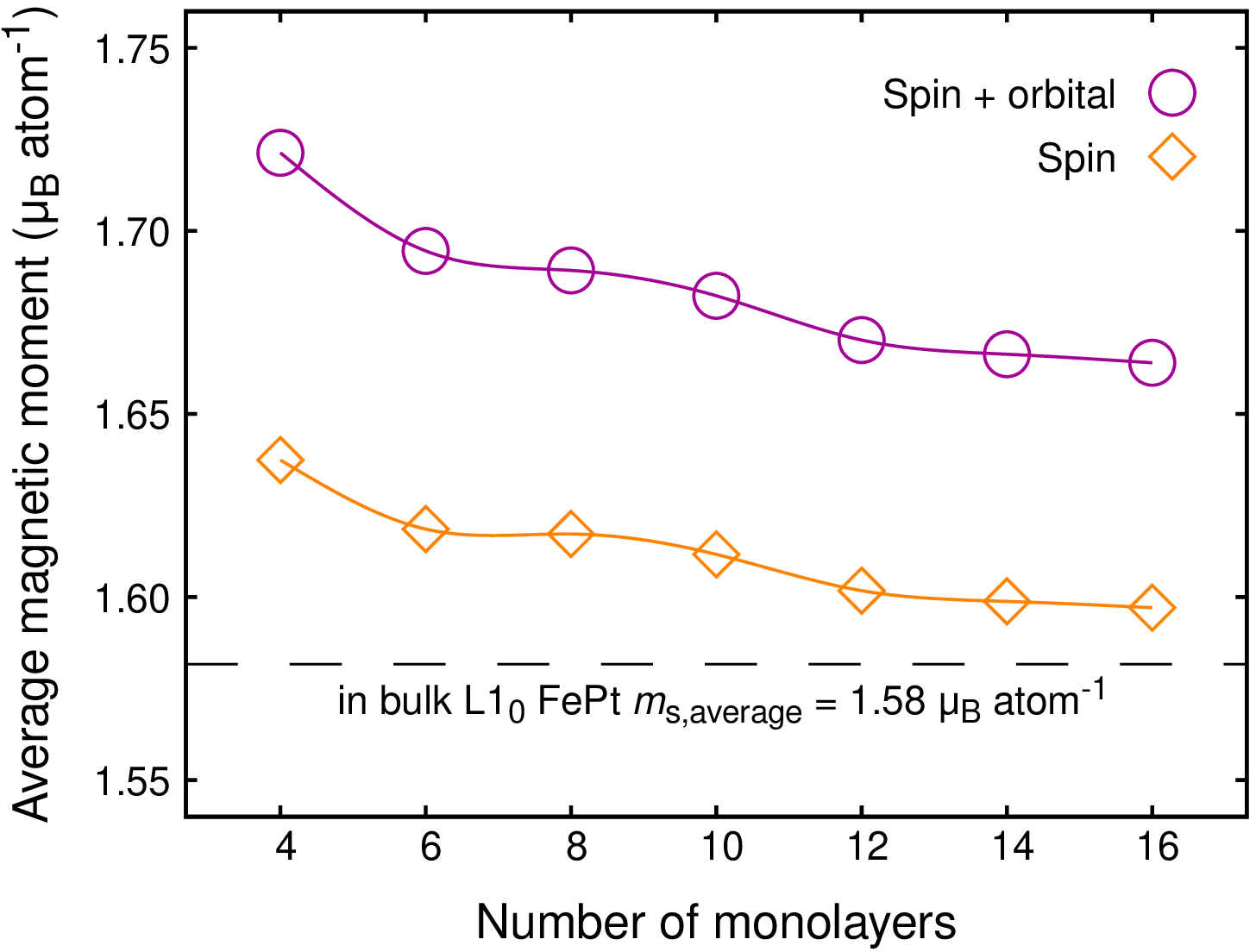}
\caption{\label{fig-mag-mom-dependence_010}
Average total (spin + orbital) and spin magnetic moments \textit{versus} the number of atomic monolayers in \lozfept{} (010) film. 
A dashed horizontal line indicates the average spin magnetic moment for the bulk \lozfept{} phase.
Calculations were performed with the PBE exchange-correlation potential using the FPLO18 code.}
\end{figure}

\begin{figure*}[t]
\includegraphics[clip,width=0.92\textwidth]{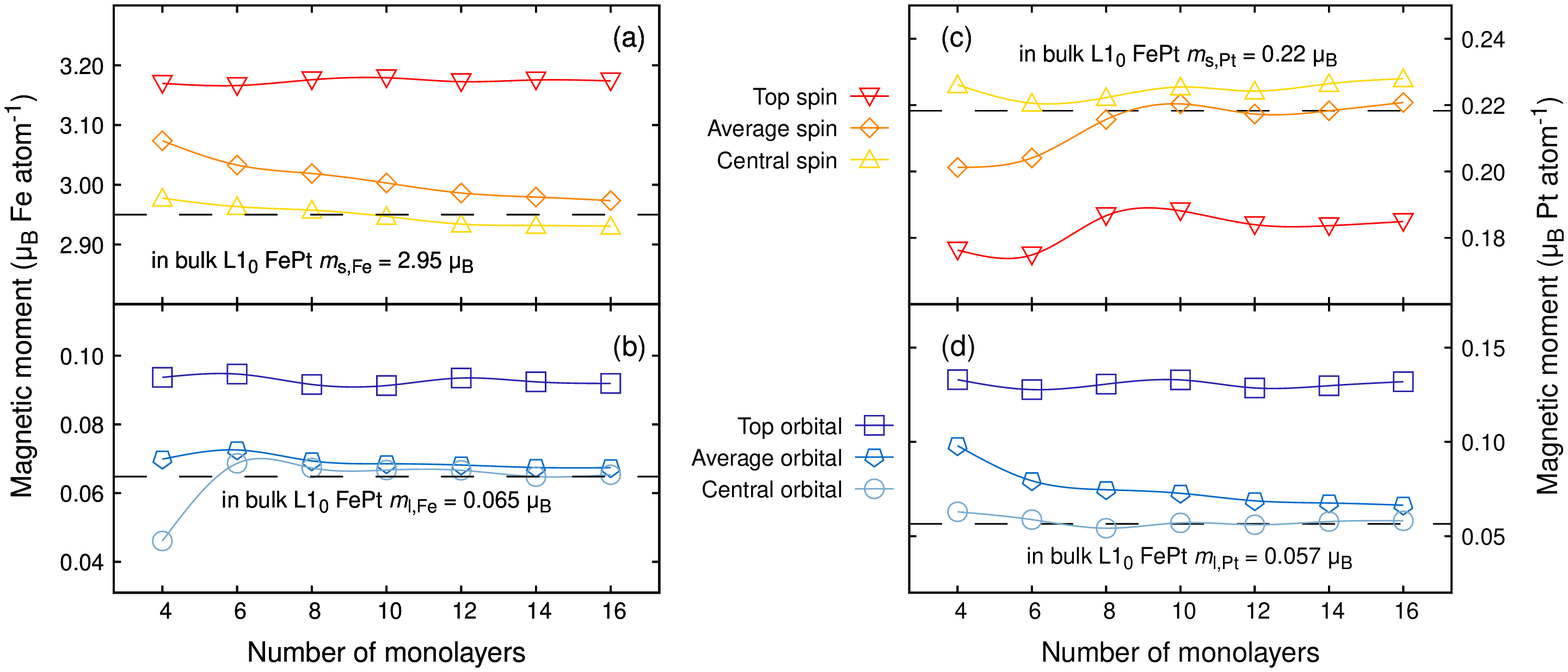}
\caption{\label{fig-mag-mom-atoms-010}
Spin and orbital magnetic moments on Fe (a, b) and Pt (c, d) atoms as a function of the number of monolayers in \lozfept{} films with (010) surface.
Magnetic moments on specified atoms for bulk \lozfept{} are indicated as dashed horizontal lines.
Calculations were performed with the PBE exchange-correlation potential using the FPLO18 code.}
\end{figure*}

%----------introduction to 010 films--------------------------
%
%
In the second part, we discuss the computational results for \lozfept{} films with a tetragonal $c$-direction oriented in the film plane, see Fig.~\ref{fig-111-structure}(b).
So far, usually on the occasion of more comprehensive studies, experimenters have observed islands/grains and areas of (100) and (010) \lozfept{} phase~\cite{hsu_situ_2000,shima_preparation_2002,ohtake_l10_2012,sepehri-amin_microstructure_2017}.
In those studies, films from 10 to 40~nm thick with areas of (100) and (010) surfaces were deposited on Ag, MgO~(001), and MgAl$_2$O$_4$ substrates.
Sepehri-Amin~\textit{et al.} found that (100) and (010) FePt grains give an in-plane component in the magnetization curves~\cite{sepehri-amin_microstructure_2017}. 
As far as we have determined, there has been no detailed study of individual FePt (100) grains and their magnetic properties.
However, intuition suggests that, at least for thicker (100) films, the magnetization should be directed in-plane and have a strong preference for alignment in the direction of the tetragonal axis.
Such a uniaxial configuration can be called in-plane fixed magnetic anisotropy.
Since the [100] and [010] tetragonal axis orientations are equivalent when FePt islands form on the substrate, the effective magnetization and in-plane anisotropy can average for larger sample areas. 
Nevertheless, we expect individual grains with (100) and (010) surfaces to exhibit strong in-plane fixed magnetic anisotropy.

%-----------structural properties of the models---------------
%
Of the two symmetrically equivalent configurations [(100) and (010)], we arbitrarily chose the one with a (010) surface for the film models we prepared.
The thickness of the 16-monolayer FePt (010) film is 3.1~nm, corresponding to a distance of 0.194~nm between atomic monolayers, which is half of the lattice parameter $a$ of the bulk \lozfept{}.
The thickness of the thinnest 4-monolayer (010) film considered is 0.78~nm.

%-----------results of magnetization direction for 10-monolayers thick film---------------%
To computationally determine the preferred magnetization direction in ultrathin \lozfept{} (010) films, for an example film with a thickness of 10 atomic monolayers, we have performed calculations of the dependence of the energy of the system on the direction of magnetization.  
We have performed calculations of the dependence of the energy of the system on the direction of magnetization to computationally
determine the preferred magnetization direction in
ultrathin \lozfept{} (010) films, for example, for a film
with a thickness of 10 atomic monolayers. 
The direction was changed according to the path of [010]~$\to$~[001]~$\to$~[100]~$\to$~[010] every 10$^{\circ}$, and the calculated results are presented in Fig.~\ref{fig-en-dir-010}.
The easy magnetization direction lies in the plane of the film and follows the direction of alternating Fe and Pt monolayers [001].
The energy difference between the easy magnetization direction and the direction normal to the film is about 1.2~meV\,atom$^{-1}$ ($\sim$13~MJ\,m$^{-3}$), which is smaller than the 1.38~meV\,atom$^{-1}$ for bulk FePt~\cite{marciniak_dft_2022}, but still very high.
The energy difference between the two orthogonal in-plane magnetization directions ([001] and [100]) is similarly very high, confirming the intuitive expectation of a strong fixed magnetic anisotropy in the plane of the film.

%---------------magnetic energy difference versus film thickness----------------%
After a complete analysis of the energy-angle dependence for the 10-monolayer film, other thicknesses were considered in the range of 4 to 16 monolayers, and the energies were calculated only for the unique magnetization directions [100], [010], and [001].
Figure~\ref{fig-en-diff-dependence} shows the calculated energy differences $E_{010} - E_{001}$ and $E_{100} - E_{001}$ for the full range of film thicknesses taken into account.
All the energy differences are positive and relatively high, which implies a high in-plane fixed magnetic anisotropy.
Except for the 10-monolayer film, the energy difference of $E_{010}-E_{001}$ is larger than $E_{100}-E_{001}$ by about 0.2~meV\,atom$^{-1}$, indicating that the magnetization direction perpendicular to the film plane has the highest energy and is, therefore, the least preferred.
In addition, the general trend shows a decrease in magnetic anisotropy with decreasing film thickness. 
However, the dependence on film thickness is not monotonic over the entire range considered, which is not surprising since many previous calculations for ultrathin films showed characteristic MAE oscillations~\cite{szunyogh_oscillatory_1997,zhang_electric-field_2009,blanco-rey_large_2021,chang_voltage-controlled_2021, cinal_magnetic_2022}.
%-----------quantum size effect - quantum well states--------------
%
Experimentally, oscillations of uniaxial magnetic anisotropy, with a period of 5.9 and 2.3 monolayers, were observed, for example, for ultrathin films of bcc Fe and fcc Co, respectively~\cite{przybylski_oscillatory_2012}.
Detailed theoretical analysis has established that the observed oscillations result from the presence of quantum-well states located in the vicinity of the Fermi level~\cite{przybylski_oscillatory_2012} and originating from the so-called quantum size effect occurring in systems of sufficiently small size, in our case films on the order of single nanometers thick.

%------------average magnetic moment versus film thicknesss-------------
%
In addition to magnetic anisotropy, the average magnetic moment of the film is an important parameter for potential applications.
Figures~\ref{fig-mag-mom-dependence_010} and \ref{fig-mag-mom-atoms-010} show the dependence of magnetic moments on film thickness.
The average total and spin magnetic moments decrease monotonically as the number of monolayers increases.
They tend to reach values for the bulk for \lozfept{} of 1.58~$\mu_\mathrm{B} \, \mathrm{atom}^{-1}$ for the spin magnetic moment and 1.64~$\mu_\mathrm{B} \, \mathrm{atom}^{-1}$ for the total magnetic moment~\cite{marciniak_dft_2022}.
The values of the average magnetic moments observed for the (010) films are slightly lower than for the (111) films (by about 0.1~$\mu_\mathrm{B}$), compare Figs.~\ref{fig-mag-mom-dependence_111} and \ref{fig-mag-mom-dependence_010}.

%-------------dependence of partial magnetic moments on film thickness-----
%
A more detailed analysis of the magnetic moments of Fe and Pt is shown in Fig.~\ref{fig-mag-mom-atoms-010}.
As the thickness of the film increases, both the average moments and the moments on the atoms in the center of the film tend to be the values determined for the bulk phase~\cite{marciniak_dft_2022}.
The magnetic moments on Fe on the surface monolayer change noticeably less than the moments on the central monolayer and the average values.
They also remain approximately constant regardless of the film thickness.
The observed non-monotonic changes in spin magnetic moments on Pt atoms are related to the variations in the energy differences presented in Fig.~\ref{fig-en-diff-dependence}.

%--------------MAE/dml decomposition--------------------------------
%

\begin{figure}[t]
\centering

\includegraphics[width=\columnwidth]{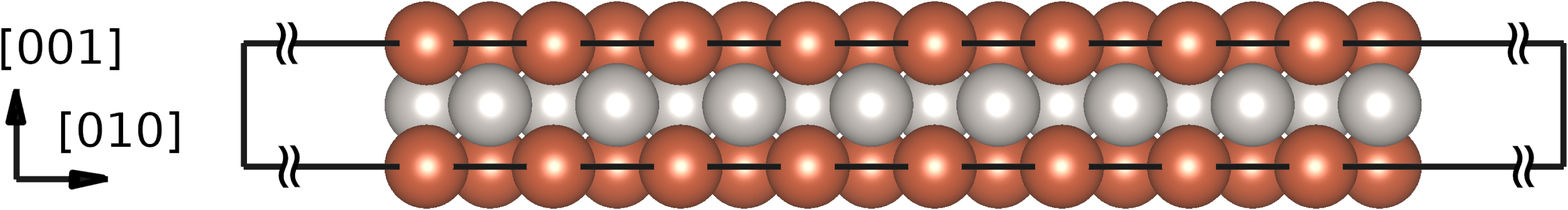}
\vspace{-1mm}

\includegraphics[trim=0 0 0 0,clip, width=\columnwidth]{fept_010_16mono_ml.eps}
\vspace{-7.5mm}

\includegraphics[trim=0 0 0 0,clip, width=\columnwidth]{fept_010_16mono_dml.eps}

\caption{\label{fig-010-dml}
Orbital magnetic moments on the atomic sites along the direction perpendicular to the film plane for 16-monolayer thick \lozfept{} (010) film calculated for [001] and [010]  magnetization directions (a).
Difference of orbital magnetic moments calculated between magnetization directions [010] and [001] (b).
Dashed lines indicate Fe and Pt contributions calculated for bulk \lozfept{}.
On top, a side view of the unit cell of 16-monolayer FePt (010) film.
The calculations were performed with the FPLO18 code with the PBE exchange-correlation potential.
}

\end{figure}

Magnetocrystalline anisotropy energy (MAE) and orbital magnetic moment are related according to the formula of Bruno~\cite{bruno_tight-binding_1989}:
\begin{equation}
\mathrm{MAE} \sim \pm \frac{\xi}{4} \Delta m_L,
\end{equation}
where $\xi$ is the spin-orbit constant,
$\Delta m_L$ is the anisotropy of the orbital magnetic moment defined as
\begin{equation}
\Delta m_L = m_L^{||} - m_L^{\perp},
\end{equation}
where $m_L^{||}$ and $m_L^{\perp}$ are orbital magnetic moments calculated in two perpendicular magnetization directions.

For bulk \lozfept{} the calculated orbital magnetic moments are 0.0648~$\mu_\mathrm{B}$ (0.0590~$\mu_\mathrm{B}$) for Fe in [001] ([110]) magnetization direction and 0.0565~$\mu_\mathrm{B}$ (0.0719~$\mu_\mathrm{B}$) for Pt in [001] ([110]) direction.
The calculated orbital magnetic moment differences are then $\Delta m_L$(Fe) = -0.0058~$\mu_\mathrm{B}$ and $\Delta m_L$(Pt) = 0.0153~$\mu_\mathrm{B}$,
which leads to the total $\Delta m_L$ equal to 0.0095~$\mu_\mathrm{B}$\,f.u.$^{-1}$.
The normalized contributions to $\Delta m_L$ from Fe and Pt are then -0.61 and 1.61, respectively. 
The above result is in qualitative agreement with the sublattice-resolved MAE calculated by Ke for \lozfept{} using perturbation theory in a tight binding approach, which gives -0.16 and 1.16 contributions for Fe and Pt (for MAE calculated as 1.25~meV\,atom$^{-1}$)~\cite{ke_intersublattice_2019}.
According to the formula of Bruno, we expect the same values of contributions for the MAE.
%, which in the case of bulk \lozfept{} we calculated as 1.38~meV\,atom$^{-1}$ (in PBE approximation).

Following the above line of reasoning, in Fig.~\ref{fig-010-dml} we present the site-resolved orbital magnetic moments and their differences calculated for two magnetization directions ([001] and [010]) for an exemplary 16-monolayer thick \lozfept{} (010) film.
We observe that in the central part of the film the values of orbital magnetic moments ($m_l$) and their differences for two perpendicular magnetization directions ($\Delta m_l$) are about the same as the values for bulk \lozfept{}.
However, the orbital magnetic moments at the film's surfaces differ significantly from the values for the bulk (similar to the previously discussed site-resolved spin magnetic moments in the FePt (111) film).
A comparable pattern emerges also for orbital magnetic moments differences, where for about five surface atomic monolayers, we observe significant deviations from the values for the central part of the film.
Assuming the applicability of Bruno's formula, we conclude that contributions of analogous proportions would also be obtained for decomposed MAE.
This indicates mainly positive dominating contributions from Pt sites and smaller negative contributions from Fe sites.
This means that, especially for films thinner than ten monolayers, modifications of individual surface monolayers may have the most significant effect on the total magnetic anisotropy of the film.

%--------------surface modifications--------------------------------
%

\begin{figure}[t]
\centering
\includegraphics[width=\columnwidth]{fept_MAE_vs_n_varous_surfaces.eps}
\vspace{1mm}

\includegraphics[width=\columnwidth]{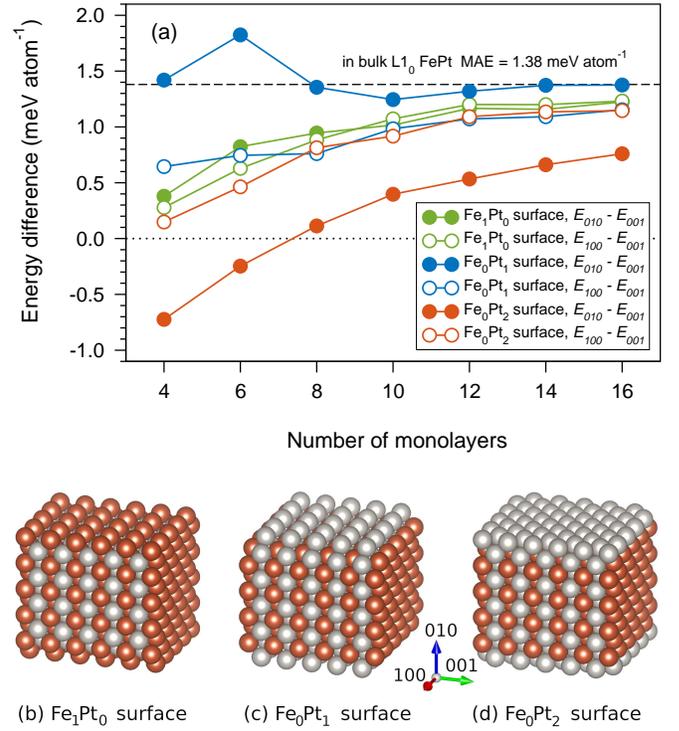}
\caption{\label{fig-surfaces_struct}
Energy differences $E_{010}-E_{001}$ and $E_{100}-E_{001}$ as a function of the thickness of \lozfept{} films with (010) surface and modified top monolayer (a).
Models of exemplary FePt (010) films of 10 monolayers thick with a modified surface single monolayer represented as unit cells multiplied several times in the directions of the film plane (b-d).
Models include surface monolayer with removed Pt atoms (a), with removed Fe atoms (b), and with Fe replaced by Pt atoms (c).
The energies $E_{001}$, $E_{010}$, and $E_{100}$ were determined for the corresponding magnetization directions using the FPLO18 with the PBE exchange-correlation potential.
}
\end{figure}

To demonstrate the impact of surface layer modification on the value of magnetic anisotropy energy, we modified one monolayer located on the surface by (1) removing Pt atoms from it, (2) removing Fe atoms from it, and (3) replacing Fe atoms with Pt atoms.
The calculations show that for all considered films with modified surfaces, the average magnetic moment per Fe atom decreases with increasing film thickness (not shown), similar to the results presented earlier for ideal films.
Figure~\ref{fig-surfaces_struct} shows our models and results of calculating the energy differences depending on the film thickness.
The presented dependencies resemble the trends observed for perfect films (without modification), see Fig.~\ref{fig-en-diff-dependence}, where we observe a tendency for the anisotropy energy to saturate with increasing film thickness.
However, significant quantitative differences are observed for magnetic anisotropy energies $E_{010}-E_{001}$ for films with surface monolayers that do not contain Fe (Fe$_0$Pt$_1$ and Fe$_0$Pt$_2$, see Fig.~\ref{fig-surfaces_struct}(c-d)).
For the Fe$_0$Pt$_1$ films, the thinnest ones do not show a substantial decrease in anisotropy energy. 
In contrast, for the Fe$_0$Pt$_2$ films, the observed energy difference values are significantly shifted toward lower ones, resulting in negative values for films with thicknesses below eight monolayers, which indicates a change in magnetization direction preference from fixed in-plane to out-of-plane.
The above results confirm that the value of magnetic anisotropy energy determined for ideal FePt films can change considerably under the influence of modification of even a single monolayer on the film surface, and that the most pronounced differences can be observed in FePt films with the smallest thickness of several monolayers.

%--------------section ending--------------------------------
%
\lozfept{} thin films and grains with the direction of the tetragonal axis aligned in the plane with respect to the substrate have great application potential and have not yet been sufficiently investigated experimentally.
It would be desirable to continue the previous studies~\cite{hsu_situ_2000,shima_preparation_2002,ohtake_l10_2012,sepehri-amin_microstructure_2017} to obtain \lozfept{} grains with the (010)/(100) surface and the high value of the ordering parameter and characterizing the magnetic properties of individual grains.
Furthermore, we also expect a similar magnetization direction anisotropy as for \lozfept{} (010) films for other L1$_0$ phases, such as FePd, FeNi, and CoPt. 
We want to emphasize that the in-plane anisotropy observed for the (010) films is unusual, as it strongly distinguishes a specific in-plane direction.
We associate the vertical alignment of the Fe and Pt monolayers relative to the film plane with books stacked on a shelf. 
Hence, perhaps this type of \loz{} film geometry could be referred to as a \textit{bookshelf} configuration, as opposed to what is usually referred to as a \textit{sandwich} geometry of a film specifying several layers arranged in a plane.

\section{Summary and conclusions}

%-----summary-------------------------------
%----------subject of study--------------------------
%
This paper presented the results of density functional theory calculations of self-standing ultrathin \lozfept{} films with (111) and (010) surfaces. 
We looked at two sets of films with thicknesses from about 0.8~nm to 3.5~nm (from 4 to 16 atomic monolayers).
%
%----------results---------------------------
%
All considered systems show a significant increase in the magnetic moment on the two near-surface atomic monolayers, leading to an increase in the average magnetic moment value relative to the value obtained for the bulk phase.
This value decreases asymptotically with increasing film thickness.
The presented dependencies of spin and orbital magnetic moments on Fe and Pt atoms on film thickness reveal additional deviations from the trends for the thinnest films regarded, with a thickness of around 1.0~nm (4-6 atomic monolayers).
Furthermore, we observed higher magnetic moment values on surfaces (111) than on surfaces (010).

%---------------111 and 010----------------------------
%
The observed increase in magnetic moments near the surface relates to the apparent charge transfer, especially in the three near-surface monolayers.
%
%--------------111----------------------------------
%
In the case of \lozfept{} (111) thin films, the alignment of the easy magnetization axis in the unique direction of the L1$_0$ tetragonal structure was experimentally confirmed earlier.
Calculations for ultrathin \lozfept{} (111) films also indicate a tilted magnetic anisotropy.
Moreover, calculations have shown that reducing the thickness of the (111) films to about a few atomic monolayers leads to a further tilt of the easy direction toward the out-of-plane.
However, an optimal angle of 45$^\circ$ from the point of view of potential applications is observed for a thickness of 6 monolayers (about 1.3~nm).

%--------------010---------------------------------%
In the case of the \lozfept{} (010) films, we showed that the easy magnetization axis is located in the plane of the film in a unique tetragonal direction, consistent with the direction of the ordering of the Fe/Pt monolayers. 
The calculations also allowed the determination of the value of the magnetic anisotropy energy in the film plane, which turned out to be only slightly lower than the value observed for bulk \lozfept{}.
This means that the \lozfept{} (010) films have an in-plane magnetic anisotropy with a very strong preference for a distinct in-plane direction, as opposed to a regular in-plane magnetic anisotropy when all magnetization directions are energetically nearly equivalent.

%----------------conclusions------------------------
%
The \lozfept{} (010) films have not yet received much attention and have found no practical application.
However, we hope that the predicted high value of in-plane fixed magnetic anisotropy will also make it possible to find a use for (010) films, e.g., as a substrate for pinning soft magnetic layers in a fixed in-plane direction.
Intuitively, we predict that analogous preferences for the directions of the easy magnetization axes can also be found in films of other \loz{} phases, such as FePd, FeNi, and CoPt.

\begin{acknowledgements}
We acknowledge the financial support of the National Science Centre Poland under the decision DEC-2018/30/E/ST3/00267.
We thank Paweł Leśniak and Daniel Depcik for compiling the scientific software and administration of the computing cluster at the Institute of Molecular Physics, Polish Academy of Sciences.
We also thank Justyna Rychły-Gruszecka and Justyn Snarski-Adamski for reading the manuscript and providing valuable comments and discussion.
We want to thank Daniel Kiphart for reading the manuscript and for his comments on the language.

\end{acknowledgements}

\bibliography{fept_l10_layers}

\end{document}